\documentclass[letterpaper, 10 pt, conference]{IEEEtran}

\usepackage{epsfig}
\usepackage{color,booktabs}
\usepackage{graphicx}
\usepackage[figuresright]{rotating}
\usepackage{amssymb}
\usepackage{amsmath}
\usepackage{setspace}
\usepackage{rotating}
\usepackage{multirow}
\usepackage{wrapfig}
\usepackage{booktabs}
\usepackage{array}
\usepackage{comment}
\usepackage{pict2e}
\usepackage{epstopdf}
\usepackage[ruled]{algorithm2e}

\usepackage{balance}
\usepackage{mathtools}
\usepackage[numbers,sort&compress]{natbib}
\usepackage{url}  %
\usepackage{subcaption}
\usepackage[font=footnotesize]{caption}
\usepackage{xparse}

\usepackage[labelformat=simple]{subcaption}

\usepackage{colortbl}

\NewDocumentCommand\jj{+u{\jj}}{\ignorespaces}

\begin{document}

\title{On-board Deep-learning-based Unmanned Aerial Vehicle Fault Cause Detection and Identification}
\author{{\bf Vidyasagar Sadhu, Saman Zonouz, Dario Pompili}\\
Department of Electrical and Computer Engineering, Rutgers University--New Brunswick, NJ, USA\\
\textit{\{hss64, saman.zonouz, pompili\}@rutgers.edu}
}

\maketitle

\thispagestyle{empty}

\begin{abstract}
With the increase in use of Unmanned Aerial Vehicles~(UAVs)/drones, it is important to detect and identify causes of failure in real time for proper recovery from a potential crash-like scenario or post incident forensics analysis. The cause of crash could be either a fault in the sensor/actuator system, a physical damage/attack, or a cyber attack on the drone's software. In this paper, we propose novel architectures based on deep Convolutional and Long Short-Term Memory Neural Networks (CNNs and LSTMs) to detect (via Autoencoder) and classify drone mis-operations based on sensor data. The proposed architectures are able to learn high-level features automatically from the raw sensor data and learn the spatial and temporal dynamics in the sensor data. 
We validate the proposed deep-learning architectures via simulations and experiments on a real drone. Empirical results show that our solution is able to detect (with over $90\%$ accuracy) and classify various types of drone mis-operations (with about 99\% accuracy (simulation data) and upto 88\% accuracy (experimental data)). 
\end{abstract}

\section{Introduction} 
\textbf{Overview and Motivation:}
The advancement in the technology of Unmanned Aerial Vehicles~(UAVs)/drones and concern of safety are pushing many government and defense organizations to use UAVs for surveillance. 
E-shopping companies like Amazon are planning to use UAVs for home delivery of their products. Further, drones are also being planned for use as mobile air-policing vehicles in some countries. The advantage that drones can replace humans in potentially dangerous situations is the main factor behind investing and researching on UAVs.

Drones or UAVs are Cyber Physical Systems~(CPS) with the increase of which, there are risks of both physical as well as cyber attacks on them~\cite{hartmann2013vulnerability}. 
Examples of cyber attacks are GPS spoofing attacks~\cite{kerns2014unmanned}, signal jamming, control command attacks, attacks on sensors~\cite{son2015rocking}, keylogging virus, etc.
Examples of physical attacks (both unintentional and intentional) are bird hits, abrupt wind changes, broken propellers, etc. 
Large sized drones/UAVs are capable of killing people if they fall from heights due to the massive potential energy possessed by them. 
The increase in use of drones/UAVs currently and in projected future makes real-time incident analysis for drones or UAVs a priority. The hobbyists owning drones/UAVs, researchers and the government all will be more curious to know the cause that prevented UAVs from not reaching its destination or deviated from its intended path. As such UAV fault/anomaly \textit{detection} as well as cause \textit{identification} are important.
Firstly it is important to detect when the UAV's operation deviates from the normal. Once it is determined that something is anomalous, more resources can be utilized to identify the cause. Identifying reasons for failure is important so that appropriate action can be taken to minimize further loss. For example, in the case of a car, knowing that the failure is caused by a flat tire will help to avoid actions such as sudden braking which will further exacerbate the situation (as it results in loss of control). Similarly in case of a flying object, for certain failures, gliding may be the best solution instead of the much obvious landing. On the other hand Artificial Intelligence~(AI) based data driven techniques are increasingly being used to solve many complex problems relating to autonomous vehicles~\cite{sadhu2019iros, Rahmati2019wuwnet, Chen2019, Sadhu2018ucomms}, smartphones~\cite{Sadhu2019percom, Sadhu2017percom, Sadhu2019tmc, Sadhu2016icac}, etc.

\begin{figure}
\begin{center}
\includegraphics[width=0.5\textwidth]{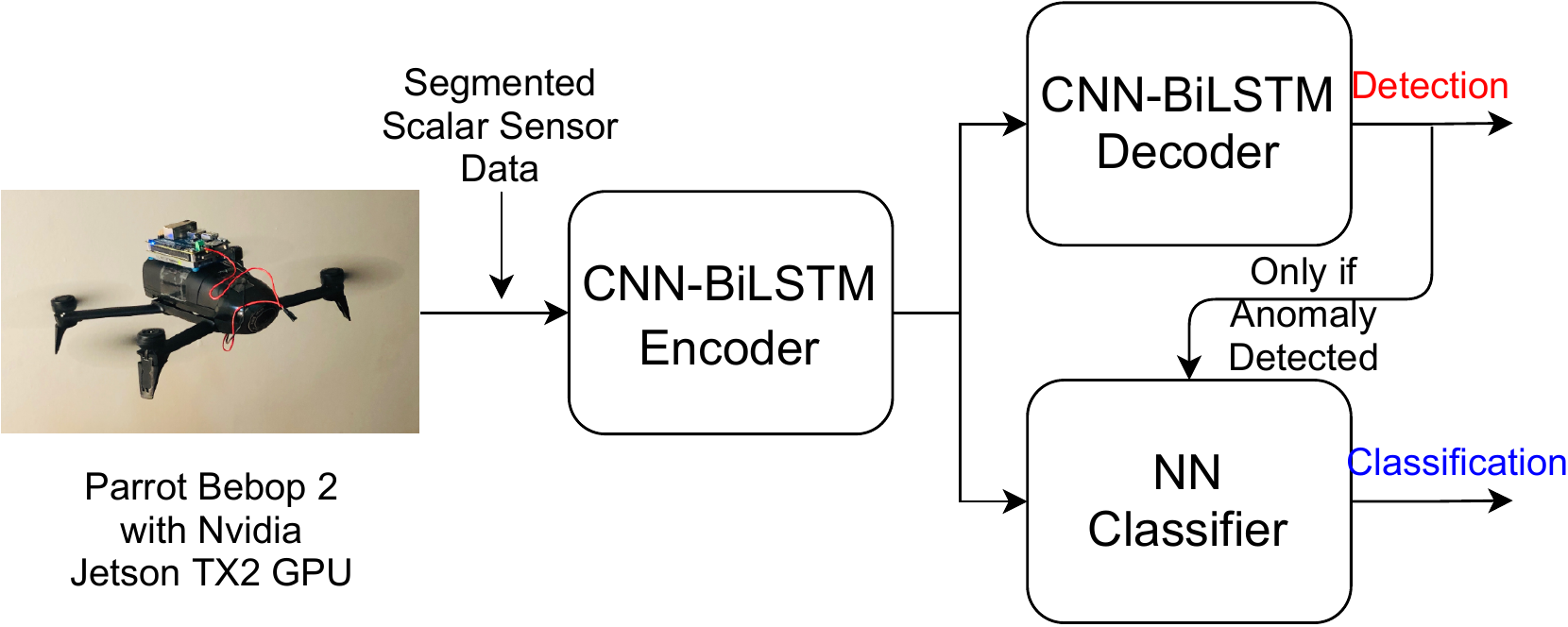}
\end{center}
\vspace{-0.1in}
\caption{An overview of our proposed on-board deep-learning based UAV fault detection and identification/classification framework.}\label{fig:overview}
\vspace{-0.25in}
\end{figure}
\textbf{Our Approach:}
Direct and continuous analysis of sensor data for real-time identification of faults is not recommended due to two reasons---(i)~it is computationally prohibitive especially on resource constrained devices such as UAVs as it requires processing huge set of sensor data; (ii)~it requires significant amount of precious on-board memory resources to store the real-time stream of sensor data. Hence, considering the resource constraints~\cite{Rahmati2019Secon, Zhao2017mass, Zhao2016, Zhao2017} of these devices,
we adopt a two-step approach (Fig.~\ref{fig:overview}) where in the identification/classification step is carried out only if an anomalous behavior is detected in the sensor data. Unlike the previous works which are mostly model-based~\cite{Hasan2019}, we follow a completely (sensor) data-driven approach (using UAV Inertial Measurement Unit~(IMU) sensor data such as accelerometer, gyroscope, etc.) for both detection and identification steps. 

The reason for choosing data-driven approach (such as deep learning techniques) over traditional model-based approaches are as follows. Deep learning techniques---(a)~have ability to learn complex patterns especially non-linear functions; the sensor data of a UAV at times of potential crash events (such as broken propeller) is highly non-linear and complex in nature; (b)~have no requirement to manually design the features from the data---the layers in a deep network learn meaningful features on their own during the training process. This also translates to another advantage of not needing domain expertise to extract the features; (c)~can work with unlabeled data in unsupervised fashion to generate features. This is very much beneficial for crash-like scenarios due to scarce availability of labelled data.
To this end, we propose a novel Convolutional Neural Network~(CNN) and Bidirectional-Long Short Term Memory~(Bi-LSTM) deep neural network based autoencoder for \textit{detection} of faults/anomalous patterns followed by a CNN-LSTM deep network for their \textit{classification/identification}.

\textbf{Our Contributions:}
The main contributions are as follows.
\begin{itemize}
\item We propose a novel Convolutional Neural Network~(CNN) and Bidirectional Long Short Term Memory~(Bi-LSTM) based deep autoencoder network architecture for real-time \textit{detection} of anomalous patterns in UAV IMU sensor data.
\item We propose a novel CNN and LSTM based deep neural network classifier for real-time \textit{identification} of the (cause of) fault/attack/crash based on the UAV IMU sensor data.
\item We induce crash scenarios by modifying the firmware internals of both the AirSim drone simulator as well as a real drone~\cite{crazyflie}.
\item We validate the proposed models via both experiments and simulations. According to the results, our solution is able to detect anomalies with over $90\%$ accuracy and can classify drone mis-operations correctly with about $99\%$ (simulation) and upto $85\%$ accuracy (experimental data).
\end{itemize}

\textbf{Paper Organization:}
In Sect.~\ref{sec:related-work}, we review related work. In Sect.~\ref{sec:prop-soln}, we describe the proposed deep-learning based UAV fault/crash detection and identification methods.
In Sect.~\ref{sec:evaluations}, we present both experimental and simulation results. Finally, in Sect.~\ref{sec:conclusions}, we conclude the paper and sketch future work.

\section{Related Work}\label{sec:related-work}

We position our work with respect to the related work that can be classified into the following categories.

\textbf{Fault Detection and Identification~(FDI):}
Much of the existing work on FDI focuses on faults in sensors/actuators in the UAV. For example, Panitsrisit et al.~\cite{elevatorfault11} propose a hardware duplication system consisting of piezoresistive sensor, pressure sensor, and current sensors to detect faults in the elevator of the UAV. Any abnormal outputs from these sensors will be detected as a failure. Rago et al.~\cite{immkalman98} present a FDI method for the failure of sensors/actuators based on Interacting Multiple-Model~(IMM) Kalman Filter approach. Actuator/sensor failures are represented by a change in the model representing the dynamics (measurements) of the system. 
Drozeski et al.~\cite{drozeski2005fault} present an FDI method using three-layer feed-forward neural network based on state information. Heredia et al.~\cite{fdi-okid-2005} uses Observer/Kalman Identification~(OKID) estimator to estimate the system state from measured input-output data. Detection of faults is done by noting deviation from the expected output beyond an accepted threshold. They use a separate estimator for each output to make the identification problem trivial. Taking a different approach, Suarez et al.~\cite{suarez2016cooperative} use kalman filtering in combination with visual techniques such as 3D projection from two observers to detect faults in the target UAV in a multi-UAV setting. 

\textbf{Anomaly Detection:}
Anomaly detection is generally an unsupervised machine learning (ML) technique due to lack of sufficient examples for the anomalous class. Within unsupervised learning, it can be broadly classified into the following categories
---statistical/regression, dimensionality reduction and distribution-based approaches.
In statistical approaches~\cite{Araya2017}, features are generally hand-made from the data such as mean, variance, entropy, etc. Certain statistical tests/formal rule checking actions are performed on these features to determine if the data is anomalous. However these approaches work only when the anomalous patterns are known apriori so they can be monitored on the sensor data. In dimensionality reduction, the data is projected onto a low-dimensional representation (such as principal components in Principal Component Analysis~(PCA)). The idea is that, this low-dimensional representation captures the most important features of the input data. Then clustering techniques such as k-means or Gaussian Mixture Models~(GMMs) are used to cluster these low-dimensional features to identify anomalies. In distribution-based approaches, the training data is fit to a distribution (such as multi-variate gaussian distribution or a mixture of them). Then given a test point, distance is calculated of this test point from the fitted distribution (e.g., Mahalanobis distance) that represents the measure of anomaly.

\textbf{Deep-learning Approaches:}
Deep-learning techniques have been widely used to solve many problems in different domains. For example, deep neural network architectures have been used to predict seizures~\cite{hosseini2016}, deep CNNs have been used extensively in content recommendation~\cite{Oord2013}, speech recognition~\cite{Sainath2015}, computer vision~\cite{Krizhevsky2012}, etc. On the other hand, RNNs and LSTMs have been used for Model Predictive Control based robotic manipulation~\cite{deepmpc15}, language modeling~\cite{Mikolov2011}, phoneme recognition~\cite{Graves2013}, etc. To the authors' best knowledge, deep learning techniques have never been used to detect/identify the cause of the UAV crashes based on sensor data. In this paper, we propose novel deep learning architectures to detect and identify the cause of UAV crashes or crash-like scenarios from drone's IMU sensor data.

\section{Proposed Solution}\label{sec:prop-soln}

In this section, we first explain our CNN Bi-LSTM autoencoder network to \textit{detect} anomalies followed by CNN-LSTM network classifier to \textit{identify} anomalies/crash scenarios.

\textbf{CNN and Bi-LSTM based Detector (`AutoEnc'):} 
Anomaly detection using unsupervised learning consists of two steps. In step~1, the system is trained with several normal examples to learn representations of the input data e.g., GMM clustering. Because we are dealing with temporal data, a sliding window approach 
needs to be adopted to learn these representations. In step~2, given a test data point, we define an anomaly score based on the learned representations, e.g., distance from the mean of the cluster.
In an LSTM autoencoder, input time series data, $\{x_0, x_1,...x_n\}$, of size $n+1$ (corresponding to one window of data segmented from full data) is fed to the encoders which consist of $n+1$ LSTM cells. The output of the last LSTM cell---called the embedding---is fed as input to a series of $n+1$ LSTM cells to generate an output, $\{a_0, a_1,...a_n\}$. The autoencoder is trained by minimizing the reconstruction error, $|\mathbf{x} - \mathbf{a}|^2$ (step~1).
\begin{figure}
\begin{center}
\includegraphics[width=3.4in]{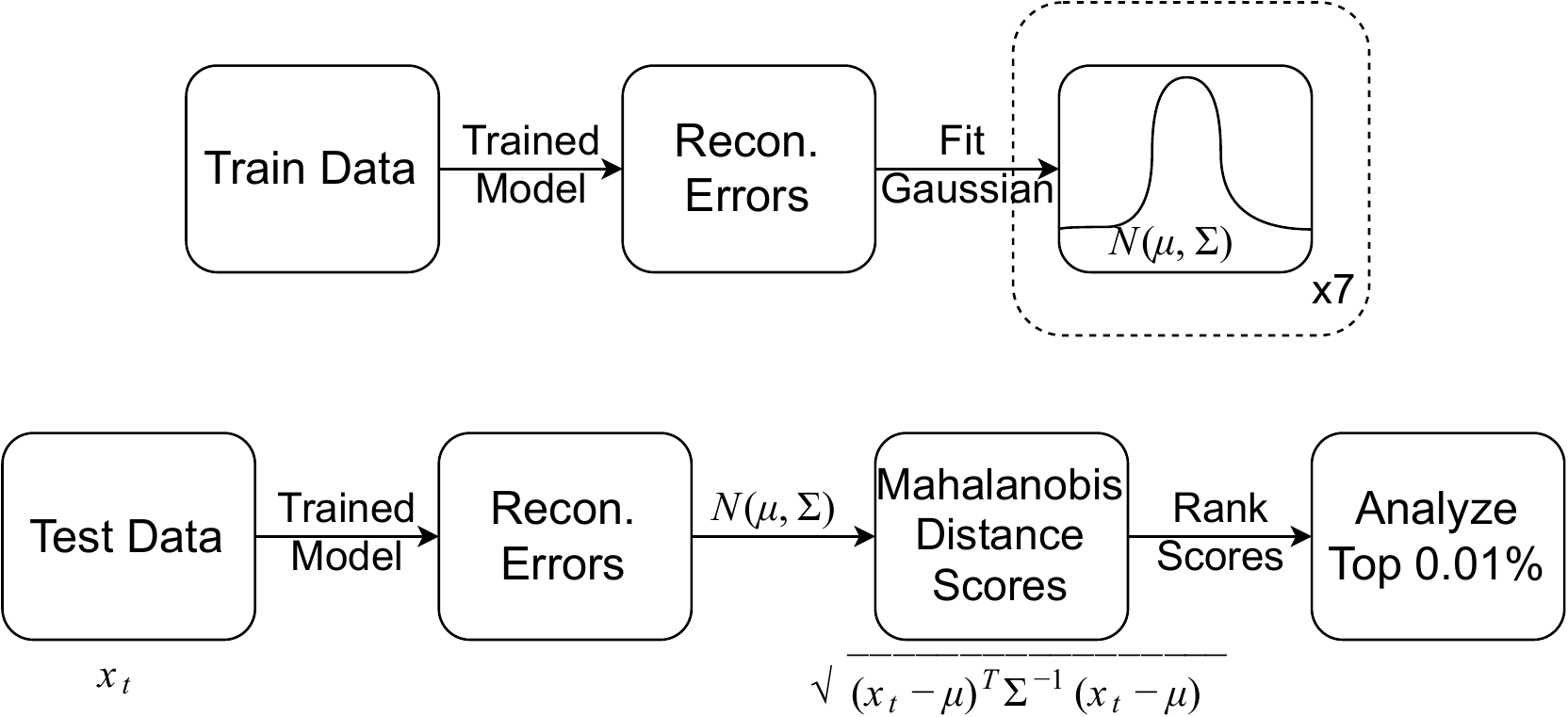}
\end{center}
\vspace{-0.15in}
\caption{(Top) After the model is trained, we fit the reconstruction errors to a multivariate gaussian model. (Bottom) Given a test data point, we first find the reconstruction error and then find the mahalanobis distance of this point w.r.t. the fitted gaussian distribution. Top scores can be analyzed as needed.}\label{fig:anom_scores_combined}
\vspace{-0.15in}
\end{figure}

\begin{figure}
\begin{center}
\includegraphics[width=3.6in]{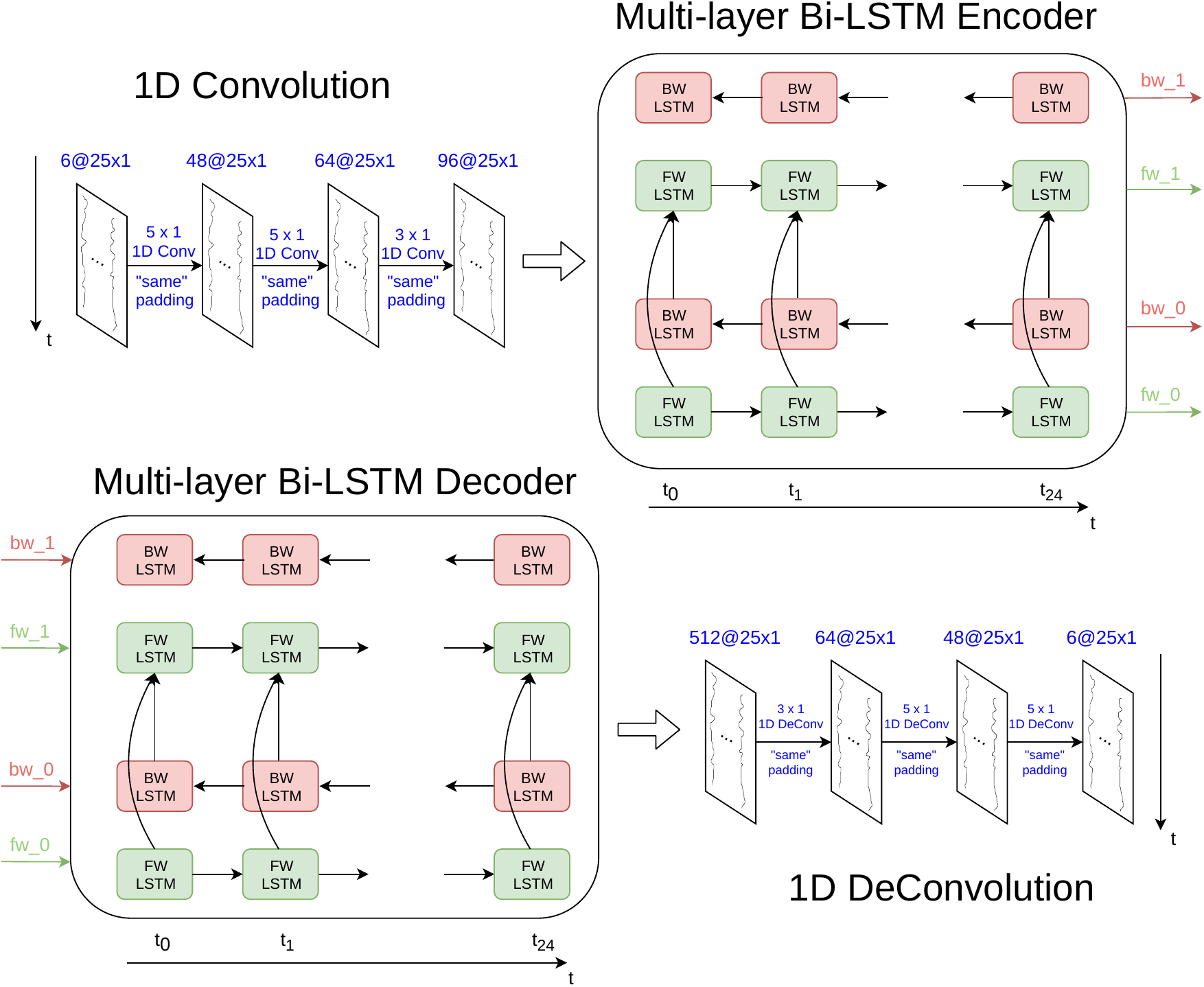}
\end{center}
\vspace{-0.15in}
\caption{Convolutional and Bi-LSTM \textit{encoder} (top) and \textit{decoder} (bottom) of the proposed autoencoder.}\label{fig:lstm_autoencoder}
\vspace{-0.15in}
\end{figure}

\textit{Convolutional Bi-LSTM Encoder:} The basic encoder in an LSTM autoencoder 
does not perform sufficiently well as it not does take into account: (i) inter channel/modal correlations (ii) directionality of data. We design an encoder that addresses these issues as shown in Fig.~\ref{fig:lstm_autoencoder}(top). It consists of a series of 1-dimensional~(1D) convolutional layers followed by bi-LSTM layers. The convolutional layers help in capturing inter-channel spatial correlations, while the LSTM layers help in capturing inter- and intra-channel temporal correlations. The number of filters, the size of filter kernel and the type of padding (with a default stride length of $1$) is indicated in Fig.~\ref{fig:lstm_autoencoder}(top). For example, in the first convolution step, $48$ filters of size $5 \times 1$ are applied to the input data of size, $25 \times 1 \times 6$ (assuming a 6-channel input data), to result in an output of size, $25 \times 1 \times 48$.  Unidirectional LSTM layers capture temporal patterns only in one-direction, while the data might exhibit interesting patterns in both directions. Hence to capture these patterns, we have a second set of LSTM cells for which the data is fed in the reverse order. Further, we have multiple layers of these bi-LSTM layers to extract more hierarchical information. All the data that has been processed through multiple convolutional and bi-LSTM layers is available in the cell states of final LSTM cells. This is the output of the encoder which will be fed as input to our decoder.

\textit{Convolutional Bi-LSTM Decoder}: The decoder performs encoder operations in reverse order so as to reconstruct the input data (Fig.~\ref{fig:lstm_autoencoder}(bottom)). It first consists of bi-LSTM layers which take the final cell states from encoder as one of the inputs (the other input being zero). 
Other input (other than the previous cell state) can be either zero or the output of the previous LSTM cell. The outputs of LSTM layers are fed as input to a series of 1D de-convolutional layers which perform reverse of convolution (also called transposed convolution) to generate data with same shape as that of input data to encoder ($6$-channel 1D data of length $25$).

Though reconstruction error can directly be used as a measure of anomaly for step~2, we design an enhanced method with further processing to obtain better results (Fig.~\ref{fig:anom_scores_combined}). After the network is trained, the train data is again fed to the trained network to capture the reconstruction errors. These errors are then fit to a multivariate gaussian distribution. Given a test data point, the reconstruction error is first calculated using the trained model. Mahalanobis distance of the error is then calculated with respect to the fitted gaussian model using the formula shown in Fig.~\ref{fig:anom_scores_combined}. These distances, which are considered anomaly scores are then sorted in decreasing order and analyzed as per requirements e.g., analyze top 0.01\%.

\begin{figure}[t!]
       \includegraphics[width=.5\textwidth]{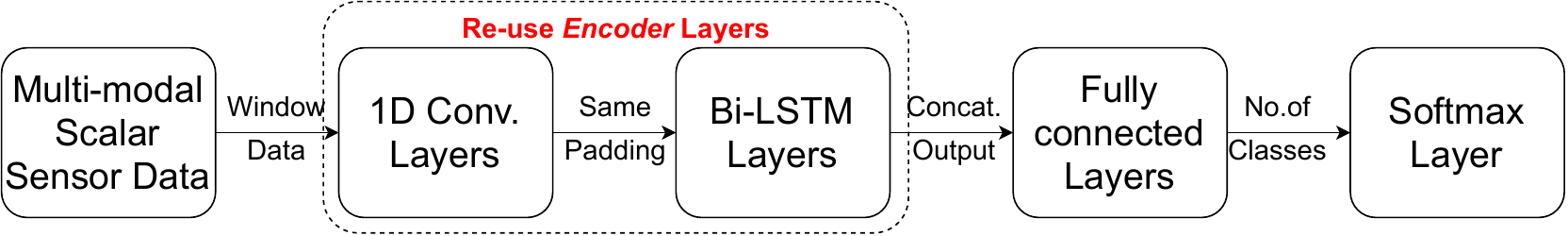}
        \caption{Proposed deep CNN and Bi-LSTM architecture for fault classification. For 1D Conv. and Bi-LSTM layers, please refer Fig.~\ref{fig:lstm_autoencoder}.}
        \label{fig:deepconvlstm}
        \vspace{-0.15in}
\end{figure}

\begin{figure}
  \centering
  \includegraphics[width=0.40\textwidth, height = 1.8in]{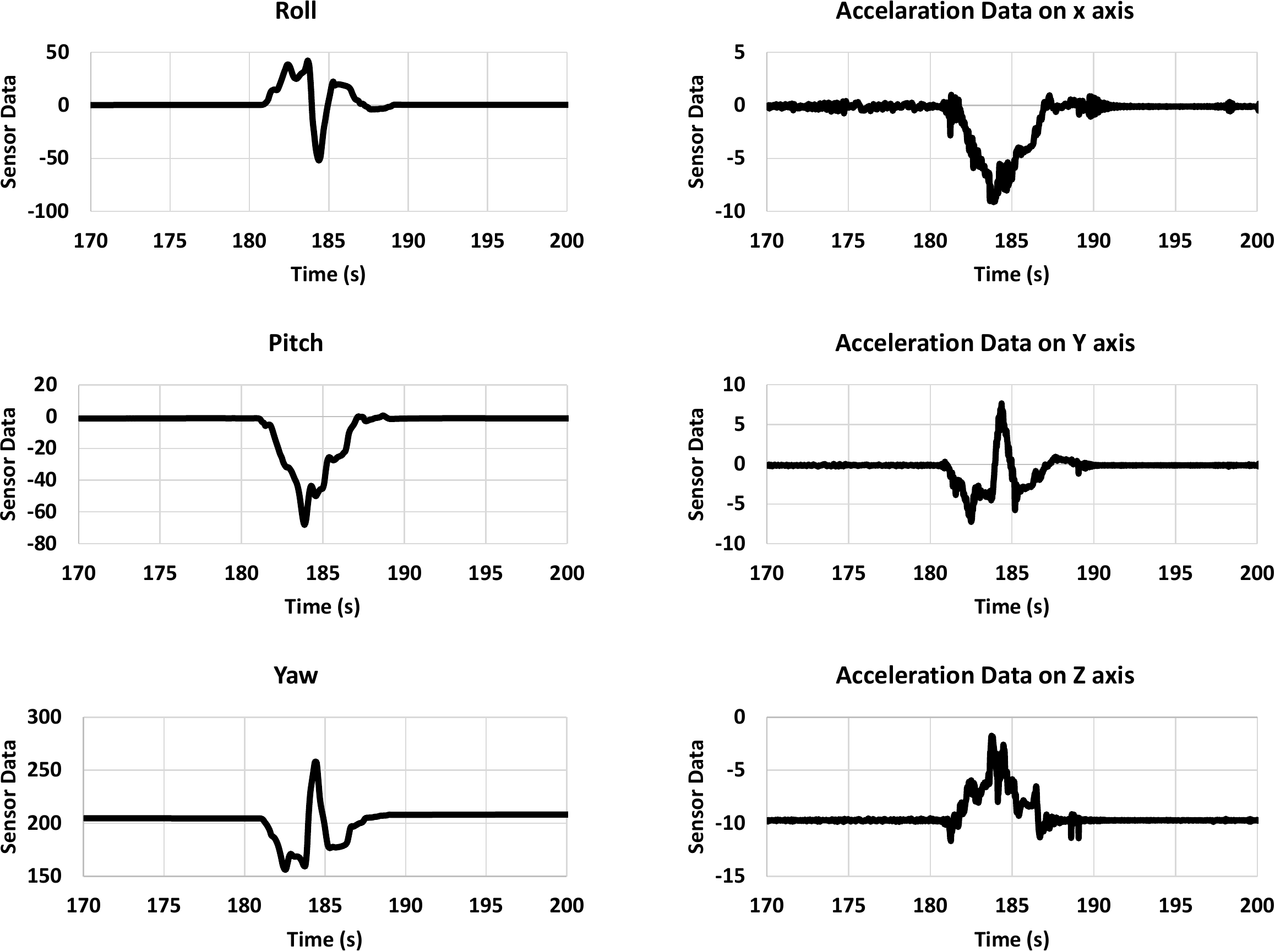}
  \caption[]{Sensor data corresponding to broken propeller scenario.}
  \label{fig:OnePropBrkn}
  \vspace{-0.2in}
\end{figure}

\textbf{CNN and LSTM based Classifier (`DCLNN'):} We consider the drone's sensor signatures in crash or crash-like scenarios to be very valuable. These signatures are mostly unique to the events that caused them. As such, we claim that these signatures can be used to identify those events by building a classifier mapping sensor signatures to events that caused them. For example, the data collected from the 3DR Solo drone~\cite{3drsolo} after a propeller was broken is shown in Fig.~\ref{fig:OnePropBrkn}. The plots show that the drone was at stable state when one of its propellers was broken, this resulted in large variations in the accelerometer and pitch-roll-yaw data (unique signatures) since the drone accelerates in a particular direction after the thrust on the broken propeller is zero. These signatures are zoomed to show the variation. 
Other sensor data such as gyroscope show similar variations. 

\begin{figure*}[th!]
        \centering 
            \begin{subfigure}[b]{0.33\textwidth}   
                \centering 
                \includegraphics[width=1\textwidth,height=1.5in]{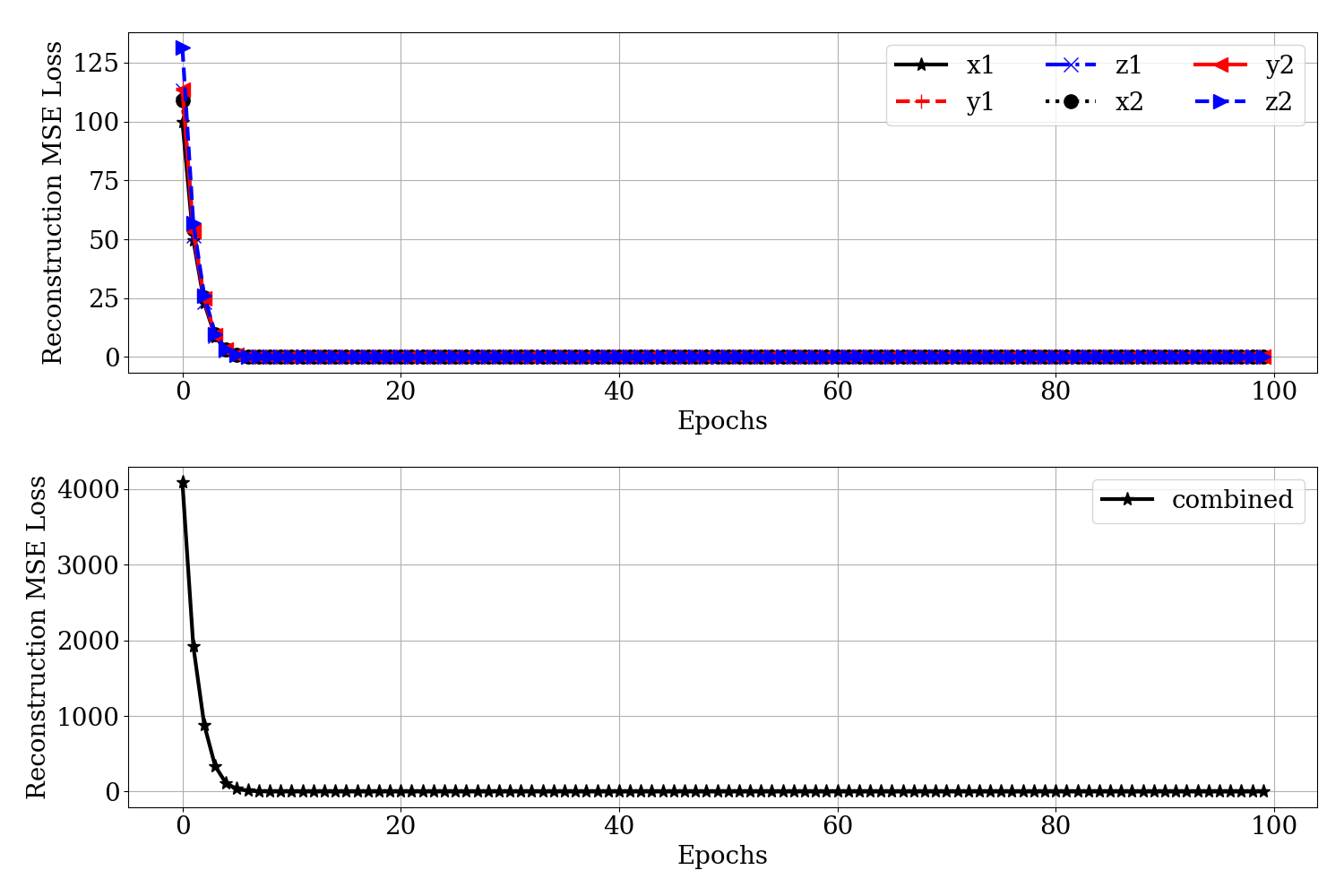}
                \caption{}
                \label{fig:detec_recon_loss}
            \end{subfigure}%
~
           \begin{subfigure}[b]{0.33\textwidth}
        		\centering
        		\includegraphics[width=1\textwidth, height=1.5in]{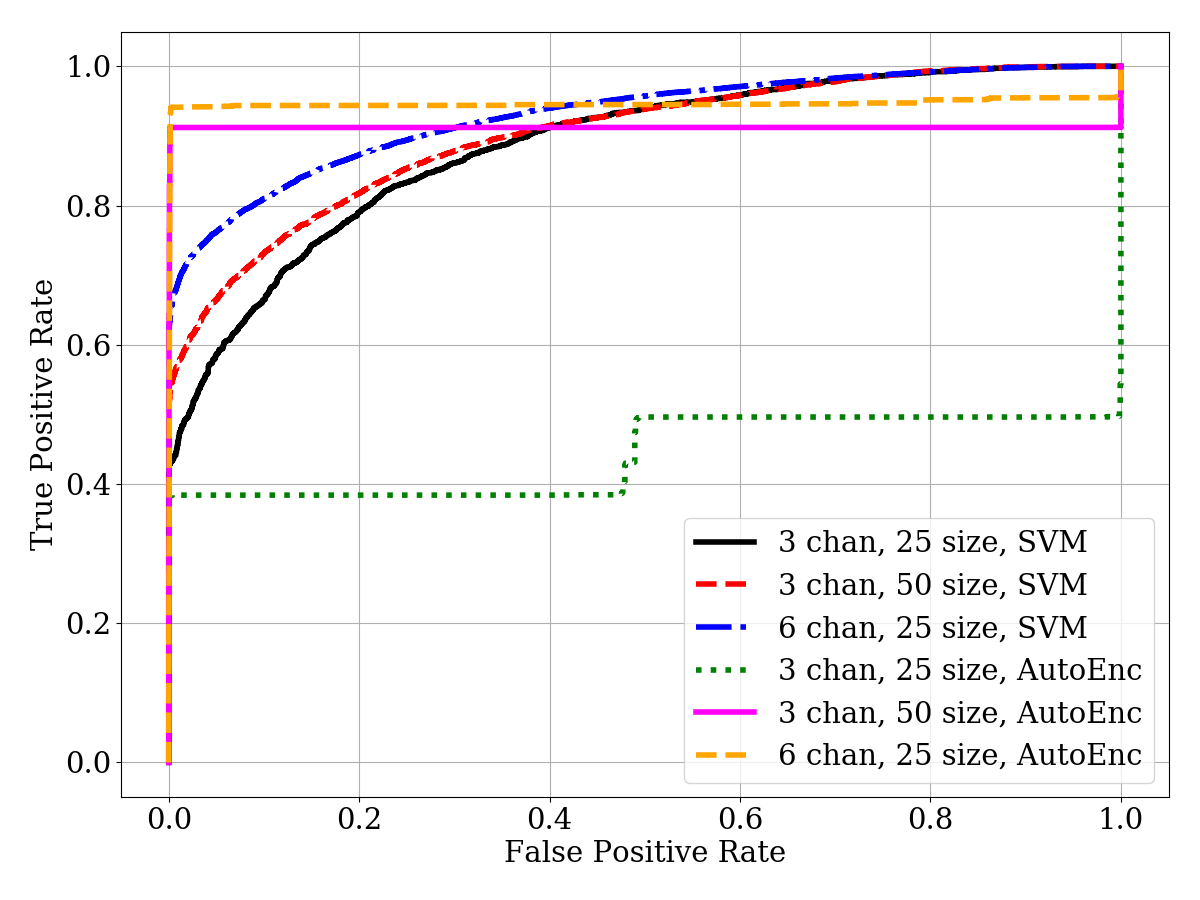}
        		\caption{}
        		\label{fig:roc_exp}
        	\end{subfigure}%
~
        \begin{subfigure}[b]{0.33\textwidth}  
            \centering 
            \includegraphics[width=1\textwidth, height=1.5in]{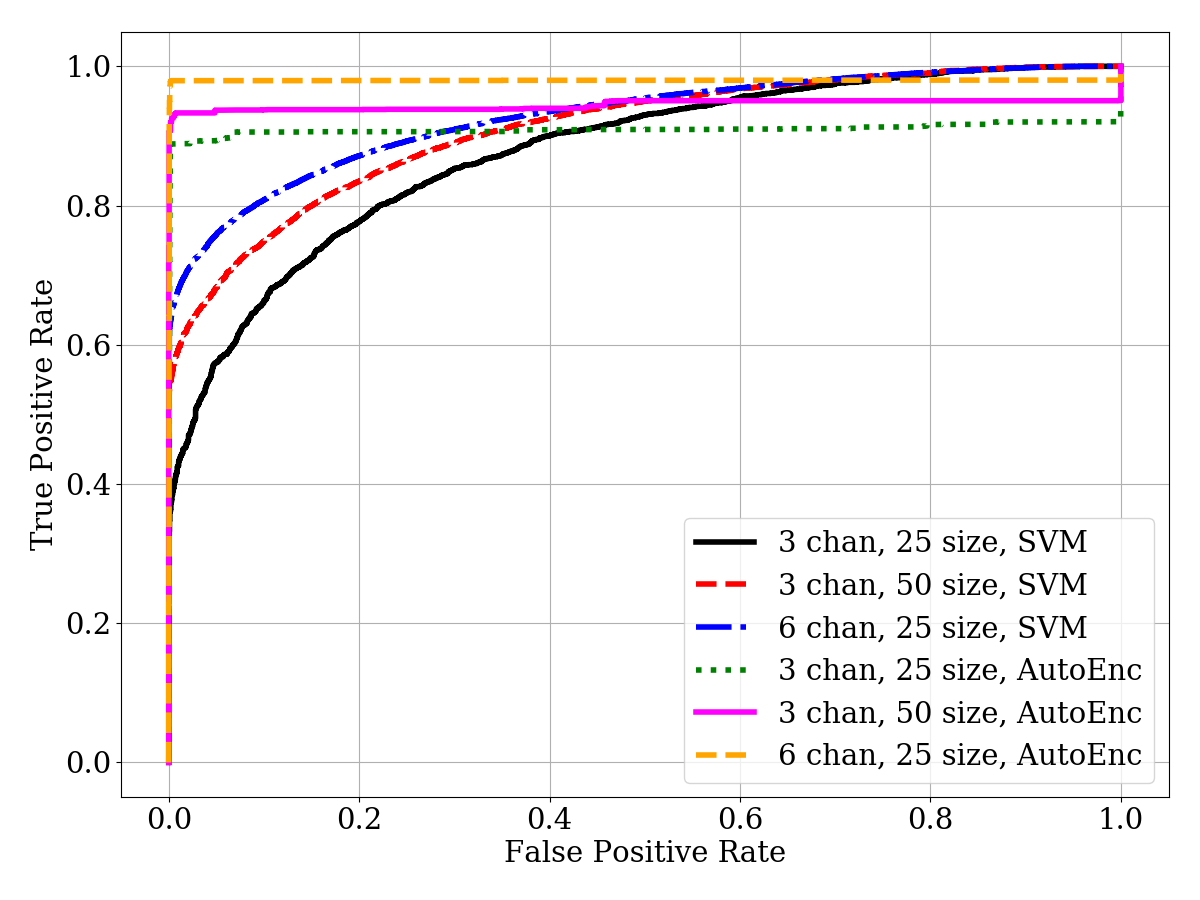}
            \caption{}
            \label{fig:roc_sim}
        \end{subfigure}
        \caption{\label{fig:app1} (a) Training reconstruction loss across each channel and combined data (experimental data); Receiver Operating Characteristic~(ROC) curve for different number of channels and window sizes---(b) experimental data; (c) simulation data.}
        \vspace{-0.15in}
\end{figure*}

We propose a novel CNN and Bi-LSTM based architecture to classify the sensor data 
in real-time to identify any potential crash scenarios. This is useful in two ways---(a)~it can be used for recovery planning to stabilize the drone using either redundant hardware or designing appropriate controller techniques that work only based on less than usual number of actuators; (b)~in cases where the crash is unavoidable, the logged sensor data can be used to identify the cause of the crash offline.
The proposed deep architecture is shown in Fig.~\ref{fig:deepconvlstm}. It consists of convolutional layers in the beginning as in the case of `AutoEnc' to extract important static/spatial features followed by LSTM layers to capture the dynamic/temporal variations in the sensor data. As such the encoder layers in Fig.~\ref{fig:lstm_autoencoder} can be reused as shown in both Fig.~\ref{fig:deepconvlstm} and Fig.~\ref{fig:overview}.
At every convolutional layer, each channel is processed by multiple kernels/filters resulting in those many feature maps in the subsequent layer. The output of the convolutional layers is time-wise unrolled and passed through bi-LSTM layers to capture the temporal dependencies in both directions. 
The concatenated outputs (of forward and backward LSTM layers) are then passed through a series of fully connected layers ending in the  
softmax layer (of length equal to number of classes) that computes the probabilities of different classes. 
We do not use pooling layers after convolutional layers as our input data has been segmented into windows and the output of the convolutional layers need to be passed through time-series LSTM layers~\cite{Ordonez2016}.
We also introduce dropout at several layers in our architectures wherein, some of the activations chosen randomly are made zero. This will act as regularization and help prevent the network to depend on some idiosyncrasies and instead learn the general structure.

\textit{Real-time operation.}
Once the models are trained offline, detection and (if necessary) identification can be done in real-time 
as it comprises only of segmenting the streaming sensor data and applying a few matrix multiplications to arrive at the result. 
Identification step is carried out only if the anomaly scores from detection step is beyond a threshold. Once the cause is identified/diagnosed in real-time, appropriate actions can be taken to stabilize/safeguard the drone operation.

\begin{figure*}[t!]
        \centering 
            \begin{subfigure}[b]{0.33\textwidth}   
                \centering 
                \includegraphics[width=1\textwidth,height=1.5in]{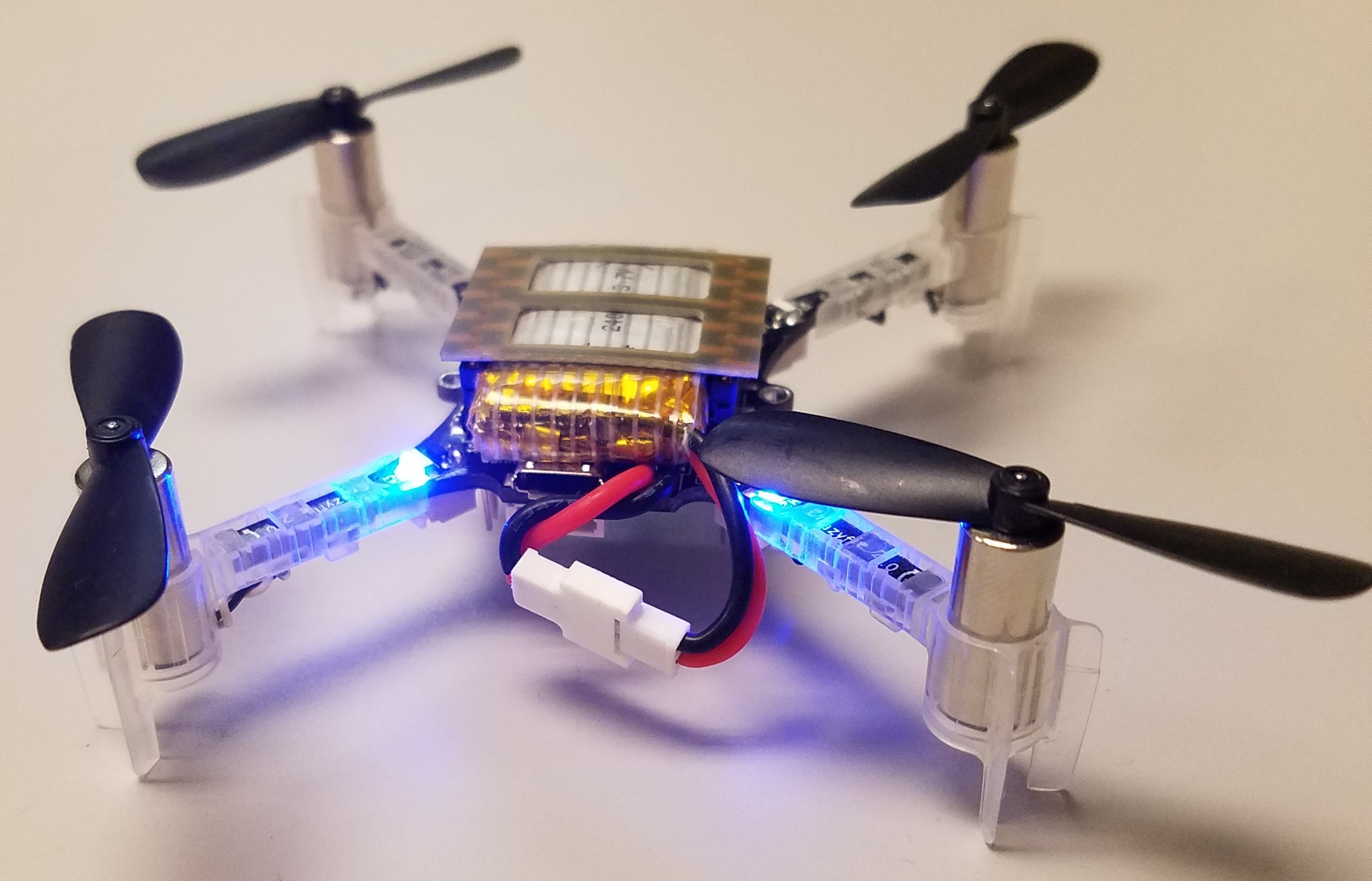}
                \caption{}
                \label{fig:crazyflie}
            \end{subfigure}%
~
           \begin{subfigure}[b]{0.33\textwidth}
        		\centering
        		\includegraphics[width=1\textwidth, height=1.5in]{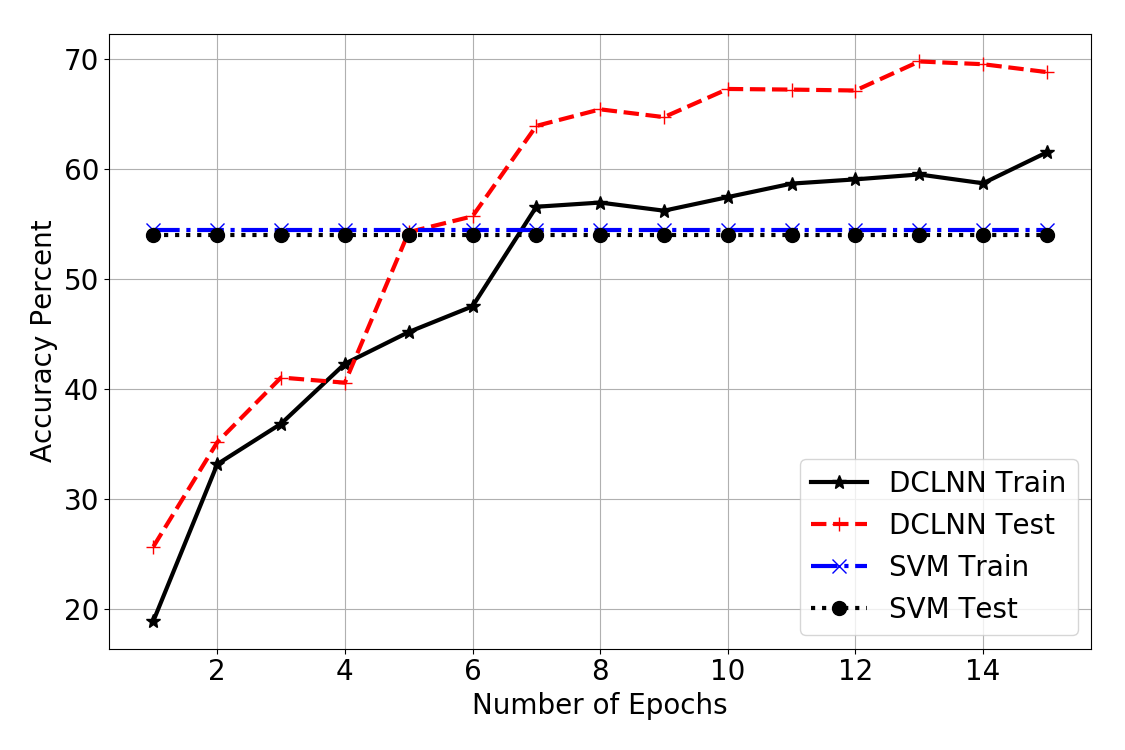}
        		\caption{}
        		\label{fig:dclnn-vs-svm-accuracy-exp}
        	\end{subfigure}%
~
        \begin{subfigure}[b]{0.33\textwidth}  
            \centering 
            \includegraphics[width=1\textwidth, height=1.5in]{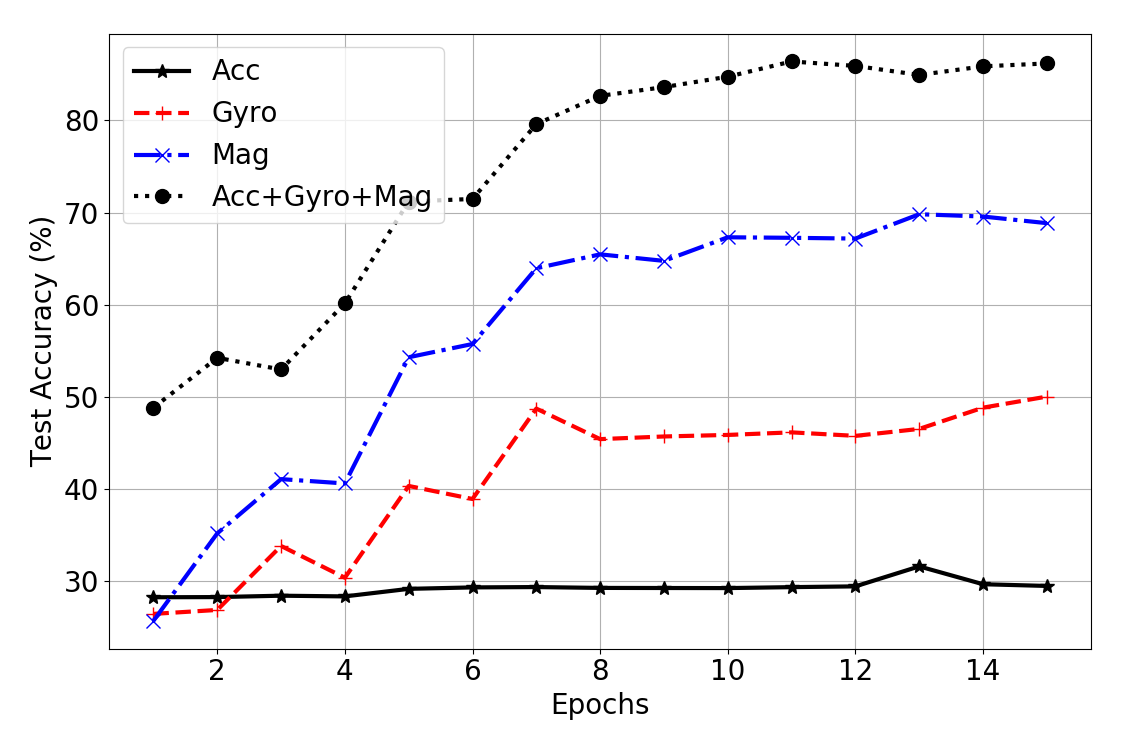}
            \caption{}
            \label{fig:test_acc_vs_chans-exp}
        \end{subfigure}
        \caption{\label{fig:app2} (a) Crazyflie 2.0 drone used for experiments. (b) Accuracy using Magnetometer data. (c) Comparison of test accuracy across different channels.}
        \vspace{-0.15in}
\end{figure*}

\section{Performance Evaluation}\label{sec:evaluations}
In this section, we first present our experimental and simulation setup followed by detection and identification results. For each case, we compare our approach with traditional machine learning classifiers such as SVM.

\textbf{Experimental Setup.}
Crash data in Fig.~\ref{fig:OnePropBrkn} is collected using 3DR Solo drone. However, we could not use the same drone for experiments as it weighs 1500 grams and easily gets damaged when it falls from heights. This is more relevant in the case of deep learning where more amount of data needs to be collected to train the models. 
For this purpose, we used another small drone, called CrazyFlie 2.0~\cite{crazyflie}, shown in Fig.~\ref{fig:crazyflie} weighing just 37 grams and much more robust to falls. 
In order to evaluate our approach, we considered a total of 15 crash scenarios (classes)---all combinations of one/two/three/four propeller breakdown cases. We modified the drone firmware by assigning a value of zero to the variables representing the propeller Revolutions Per Minute~(RPM) at appropriate levels within the software to induce crash. However, due to firmware limitations, we could not successfully induce the following cases---all four cases of three-propeller breakdown and 2 cases of diagonal two-propeller breakdown, resulting in only 9 classes totally. We collected the data for 30 runs; in each run, the drone would have a 2 second upflight time, an 8 seconds of hovering time followed by crash corresponding to one of the 9 classes. We collected Accelerometer, Gyroscope and Magnetometer data sampled at 100 Hz (maximum supported rate) for each crash event.

\textbf{Training Description.}
We used $70\%$ of data for training and $30\%$ for testing in both experiments and simulations. Since our data is time-series, we processed it into windows of size $100, 50, 25$ timesteps with a stride length of $10$ for suitable analysis. All the data till the crash is used for training/testing AutoEnc as it considered normal operation. Transition data at the time of crash is used for training/testing DCLNN along with corresponding class label (i.e., crash scenario mentioned above).
We used tensorflow to build, train and test our models with a minibatch size of $512$ windows. The number and the size of filters used for CNN layers is as shown in Fig.~\ref{fig:lstm_autoencoder}. 
We used two-layers of bi-LSTMs with a hidden size of $256$ units for each LSTM cell. 
We trained the overall network for $100-300$ epochs using Adam optimizer~\cite{Kingma2015} with a learning rate of $0.01$ and epsilon value of $0.01$.

\textbf{Detection Results.} Fig.~\ref{fig:detec_recon_loss} shows the training reconstruction loss of the experimental data for our AutoEnc model over several training epochs for 6-channels (accelerometer and gyroscope) individually and in combined form. We can notice that as the model learns, the reconstruction loss decreases and converges. We considered detection as a binary problem. Fig.~\ref{fig:roc_exp} shows the Receiver Operating Characteristic~(ROC) curve for the experimental data compared with SVM classifier (with best parameters: Radial Basis Function~(RBF) kernel, $C=10$, $\gamma=0.1$). We can notice that: (i) accuracy increases as the number of channels or the window size considered is increased; (ii) AutoEnc performs better than SVM except for the 3-channel, 25 window size case. We believe the reason is due to lack of sufficient data for training AutoEnc (as AutoEnc is a deep learning network more data is required to train the model). For the same reason, this behavior is not observed in 6-channel case. Fig.~\ref{fig:roc_sim} shows the ROC curve results for our simulation data (we describe the simulator setup and data collection below). We can observe accuracy greater than $97\%$ and also notice that we do not observe the similar problem observed in experimental data.

\begin{figure*}[t!]
        \centering   
           \begin{subfigure}[b]{0.26\textwidth}
        		\centering
        		\includegraphics[width=1\textwidth, height=1.5in]{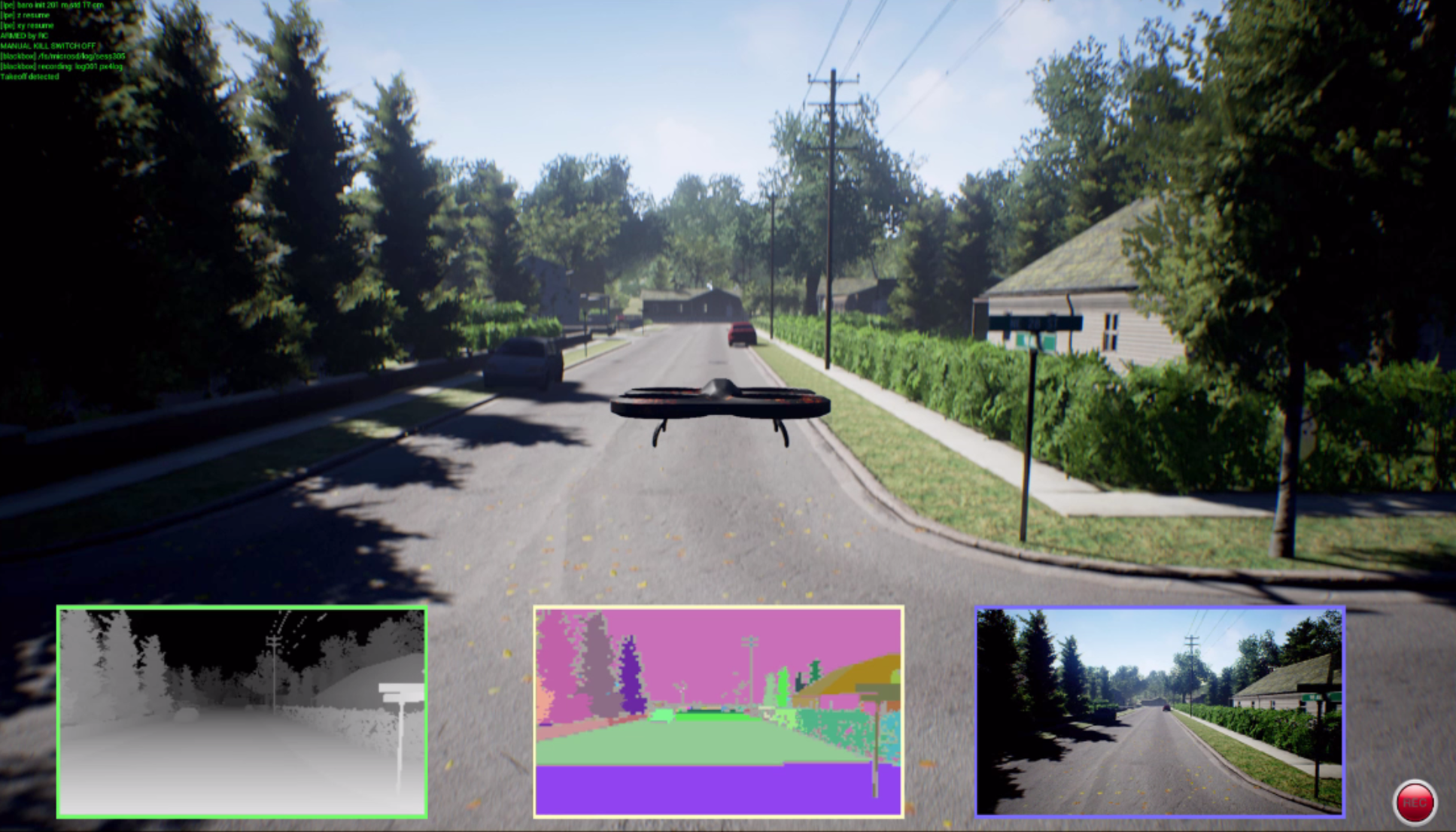}
        		\caption{}
        		\label{fig:drone_depth_materials}
        	\end{subfigure}%
~
        \begin{subfigure}[b]{0.47\textwidth}  
            \centering 
            \includegraphics[width=1\textwidth]{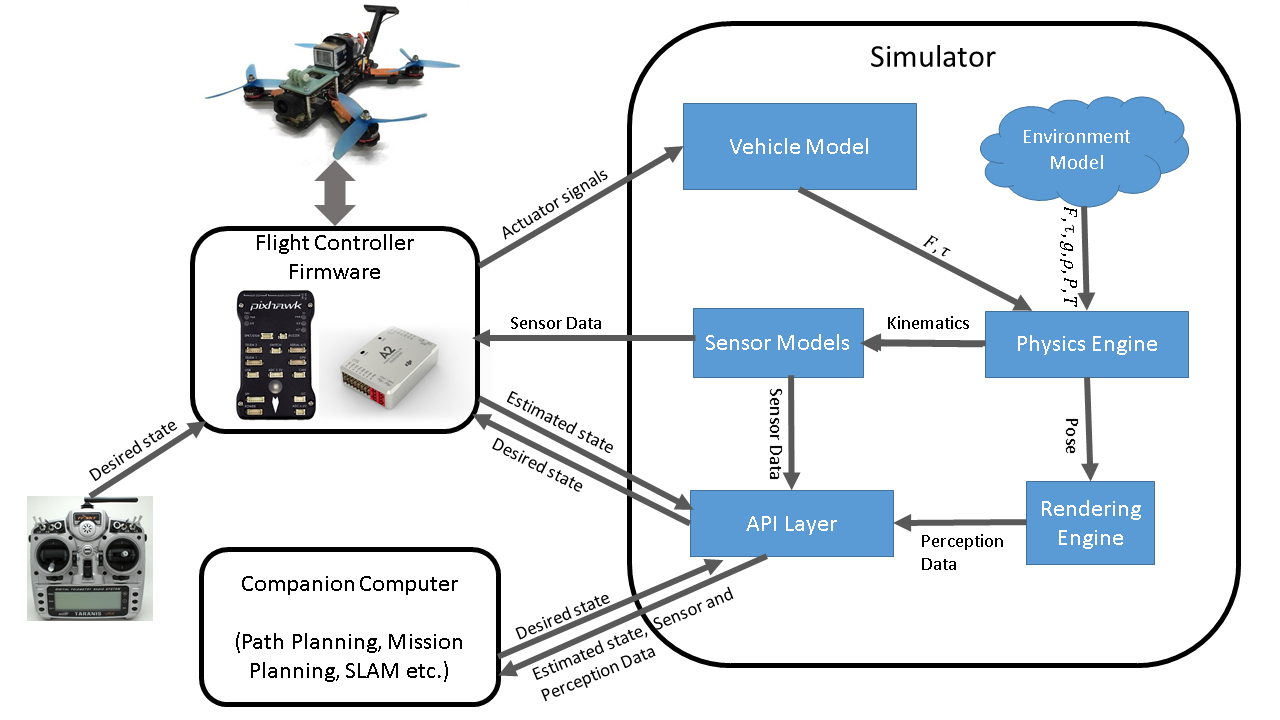}
            \caption{}
            \label{fig:arch}
        \end{subfigure}%
~
        \hspace{-0.1in}
        \begin{subfigure}[b]{0.26\textwidth}   
            \centering 
            \includegraphics[width=1\textwidth,height=1.5in]{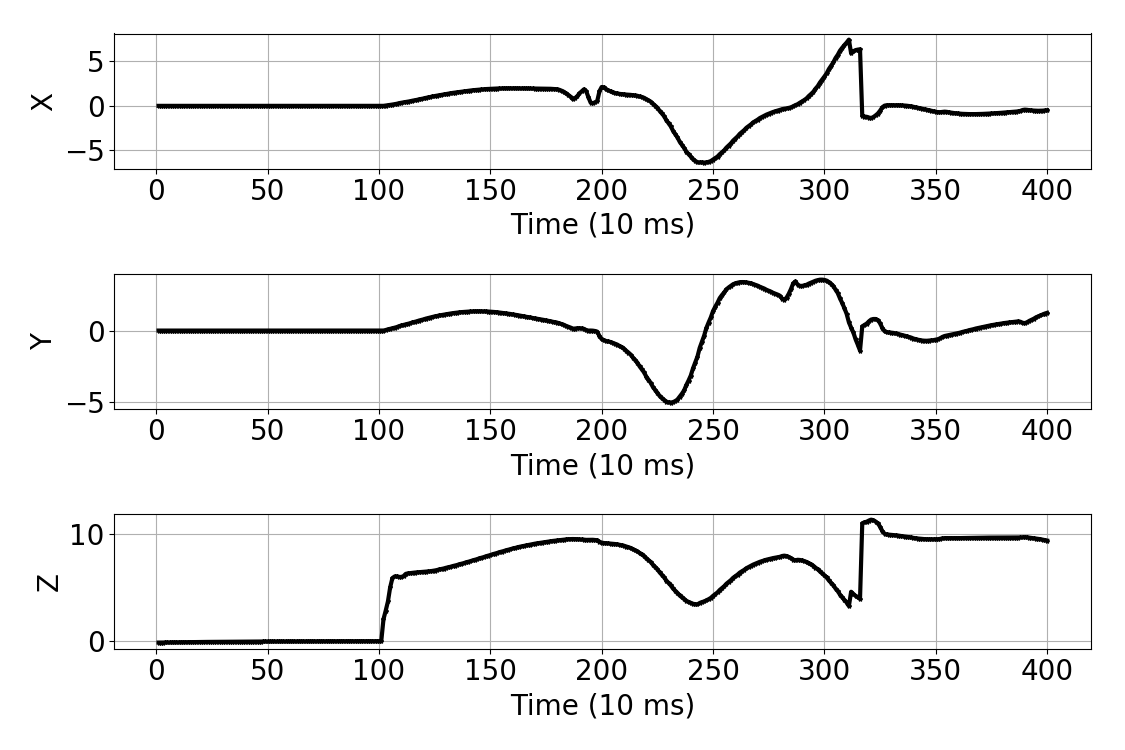}
            \caption{}
            \label{fig:crash_1_prop}
        \end{subfigure}
        \caption{\label{fig:app3} (a) A snapshot from AirSim shows an aerial vehicle flying in an urban environment. The inset shows depth, object segmentation and front camera streams generated in real time. (b) The architecture of the system that depicts the core components and their interactions. (c) Three-dimensional accelerometer data after one of the propeller's RPM is made zero (simulating broken propeller crash).}
\end{figure*}

\textit{As can be observed from these experimental results, the detection accuracy can still be improved. We believe the reason for this is the unstable nature of our drone (Crazyflie 2.0) used for experiments.} As the drone is very light weight (only 37 grams) and hard to control, it exhibits a certain element of randomness. This is also reflected in crash signatures making it hard for the network to learn meaningful representations as we show below for experimental data. On the other hand, we are also restricted in using larger drones due to---(i) state laws prohibiting flying drones outdoors without an expensive license and it is very risky to fly these drones indoors; (ii) these drones get easily damaged when they fall from heights making them unsuitable for crash experiments. \textit{To find a sweet spot between these two extremes, we decided to use a realistic drone simulator to circumvent these problems}. We felt the simulation should not be simplistic, ignoring the physical aspects including environmental objects such as trees and poles, kinematics such as drag, fiction and gravity, etc. For these reasons, we adopted Microsoft's AirSim drone simulator~\cite{MicrosoftResearch} which is an open-source simulator written in C++. The simulator tries to simulate the real-world as closely as possible by trying to include all effects involved including collisions (see Fig.~\ref{fig:drone_depth_materials} for a screenshot of the simulator).

\textbf{Simulator Setup.}
We first describe the simulator architecture followed by data collection method.

\underline{Simulator Architecture:}
The core components of AirSim include environment model, vehicle model, physics engine, sensor models, rendering interface, public API layer and an interface layer for vehicle firmware as depicted in Figure~\ref{fig:arch}. It is necessary to have simulated environments have reasonable details. For this purpose, AirSim leverages rendering technologies implemented by Unreal engine~\cite{Karis2013}. In addition, AirSim also utilizes the underlying pipeline in the Unreal engine to detect collisions. The AirSim code base is implemented as a plugin for the Unreal engine. For in-depth details on different \textit{realistic} models used in the simulator, please refer~\cite{MicrosoftResearch}.

\underline{Inducing Crash:}
In order to simulate a broken propeller, we make its RPM to zero by supplying zero current to its motor. However, for this to work successfully, it is necessary to constantly provide zero current to the effected motor. A one-time operation would not be sufficient as the currents to the motors are generated in a high-frequency update loop using a PID controller and hence correct current values (which do not induce crash) are provided to the motors in subsequent update loops. Hence, we modified the firmware code to make the effected propeller's motor current to zero inside the update loop itself so that its RPM is continuously zero, resulting in a crash. Fig.~\ref{fig:crash_1_prop} shows the accelerometer data (x,y,z axes) after the crash is induced by making the RPM of one of the propellers to zero. By comparing with the actual drone crash data (accelerometer) from 3DR Solo drone in Fig.~\ref{fig:OnePropBrkn}, we can see that the simulation data is more complex and hence difficult to learn than the former. We successfully simulated all 15 crash scenarios (classes). In all these scenarios, we collected 18-channel data viz., linear and angular versions of acceleration, velocity, and position along all the three axes from the start of the crash until the end. We repeated this experiment $300$ times to account for noises and gather sufficient data. We have included the video demo of the crash experiments run in AirSim simulator along with this submission. For results below, we used only 3-channel linear acceleration data (instead of all 18 channels), unless otherwise specified, as there is a need to limit the amount of data to process on a real drone.

\textbf{Identification/Classification Results. }
We used our DCLNN architecture in Fig.~\ref{fig:deepconvlstm} with three convolutional layers ($[48,64,96]$ filters with kernel sizes $[5,5,3]$) followed by two bi-directional LSTM layers ($128$ units) and one dense/softmax layer. The training is performed for 15 epochs in mini-batches of size $64$ using Adam optimizer and categorical cross entropy as loss function. We now present identification results using experimental data.

\begin{figure*}[t!]
        \centering 
            \begin{subfigure}[b]{0.33\textwidth}   
                \centering 
                \includegraphics[width=1\textwidth,height=1.5in]{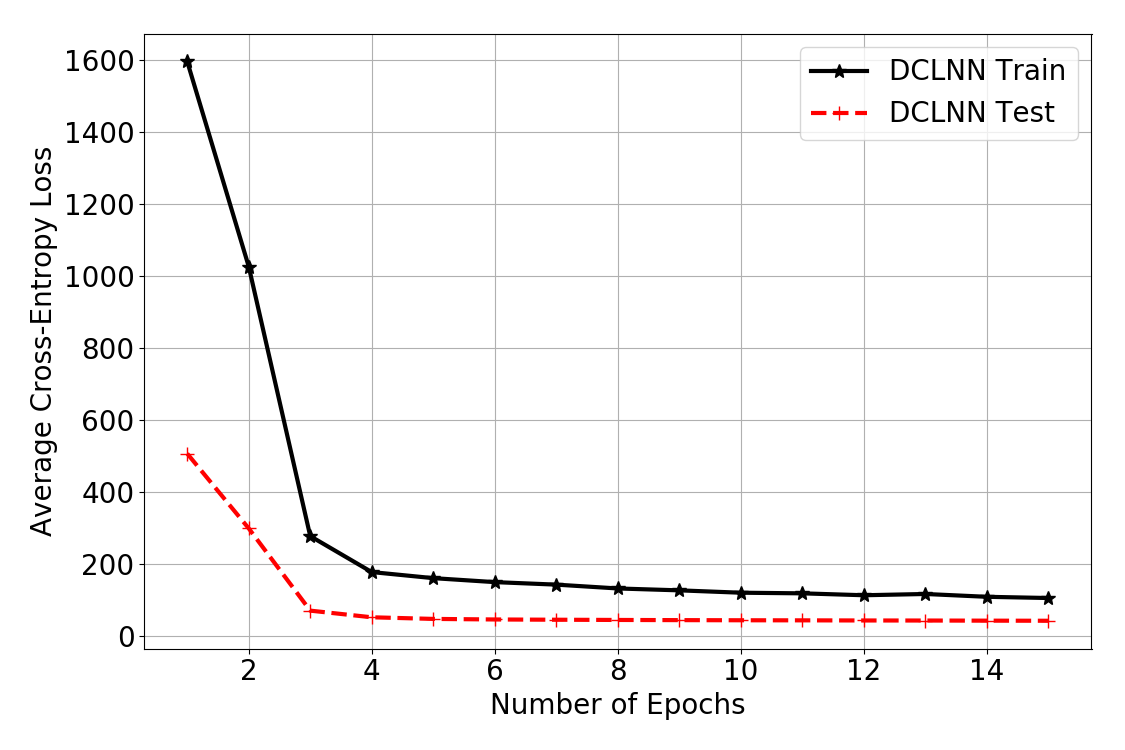}
                \caption{}
                \label{fig:loss-vs-epochs}
            \end{subfigure}%
~
           \begin{subfigure}[b]{0.33\textwidth}
        		\centering
        		\includegraphics[width=1\textwidth, height=1.5in]{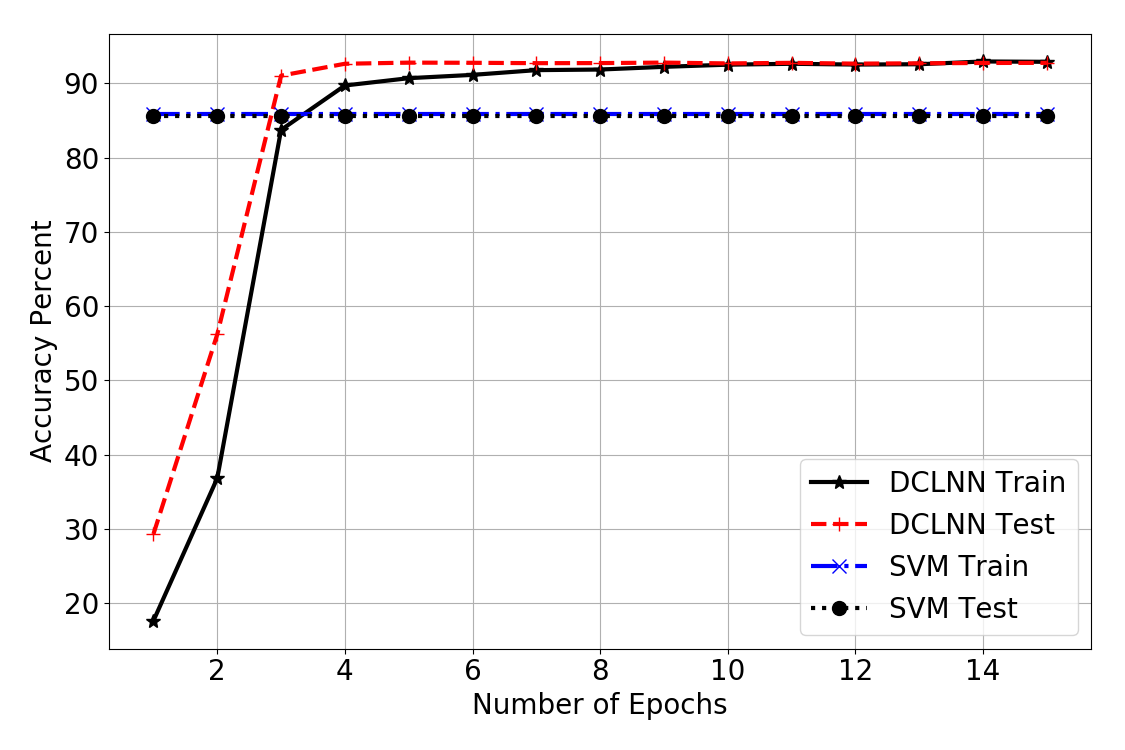}
        		\caption{}
        		\label{fig:dclnn-vs-svm-accuracy}
        	\end{subfigure}%
~
        \begin{subfigure}[b]{0.33\textwidth}  
            \centering 
            \includegraphics[width=1\textwidth, height=1.5in]{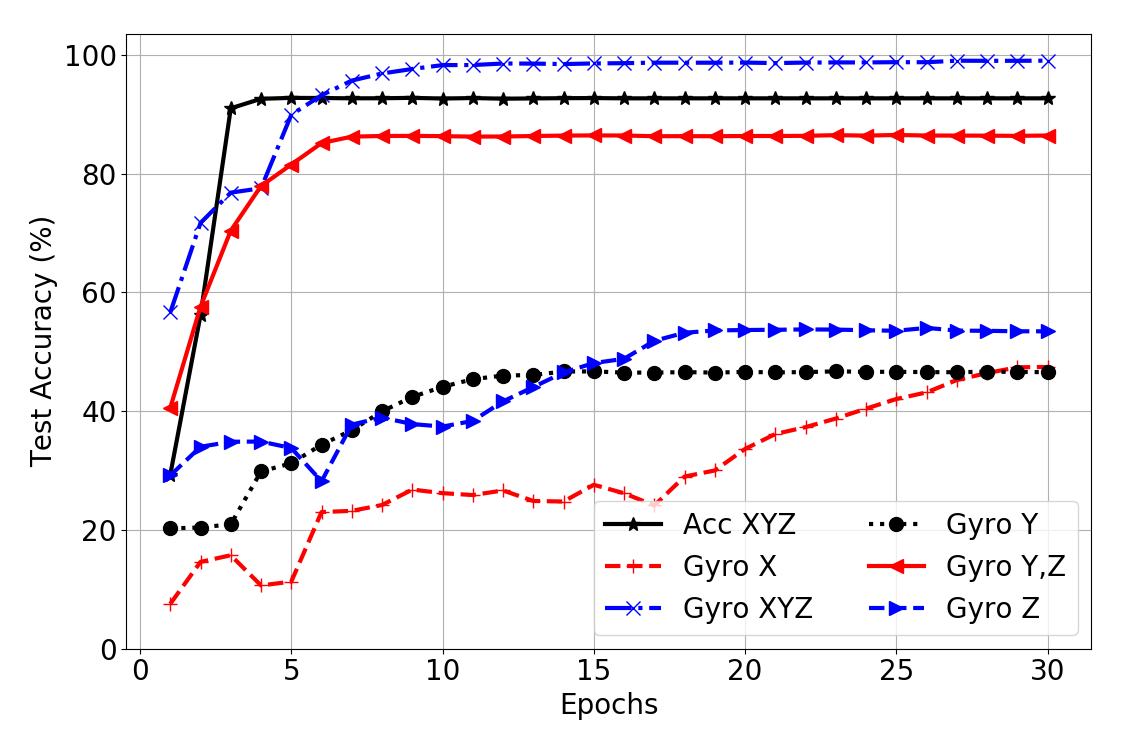}
            \caption{}
            \label{fig:test_acc_vs_chans}
        \end{subfigure}
        \caption{\label{fig:app4} (a) Cross-entropy loss vs. number of training epochs for both train data and test data. (b) Accuracy vs. number of training epochs compared between DCLNN (ours) and SVM classifier for both train and test data. (c) Accuracy on test data when the channel data fed to the network is varied.}
        \vspace{-0.15in}
\end{figure*}

\underline{Comparison with SVM.} Fig.~\ref{fig:dclnn-vs-svm-accuracy-exp} shows the accuracy of our model (DCLNN) compared against classical machine learning classifier such as Support Vector Machine~(SVM). Accuracy is defined as the percentage of windows classified correctly. We can notice that DCLNN's accuracy increases as the training is carried out over several epochs (cross-entropy loss also reduces accordingly but is not plotted due to space limitations), reaching to $70\%$ finally. We can also notice that it performs better than SVM classifier (RBF kernel, $C=10$, $\gamma = 0.01$), which could only achieve $54\%$ accuracy. Due to space limitations, we have shown comparison only with Magnetometer data; in other channels too, DCLNN performs better than SVM in a similar manner.

\underline{Variation with Channels.} There is a need to limit the amount of data processed during inference considering the resource scarcity of UAVs. We were curious if all the channels contributed equally to the learning of the network. For this purpose, we trained our network by giving different channel data each time and plotting accuracy on test data as shown in Fig.~\ref{fig:test_acc_vs_chans-exp}. We can notice that in 3-channel scenario, Magnetometer performs best with $70\%$ accuracy, while combining all 9-channel data yields an accuracy of about $85\%$. 
We now present results corresponding to simulation data.

\underline{Comparison with SVM.} Fig.~\ref{fig:loss-vs-epochs} shows the training and testing loss as our network is trained over several epochs. We notice that the loss reduces and converges to a value. This indicates that the network is learning over time and fits the data reasonably. Fig.~\ref{fig:dclnn-vs-svm-accuracy} shows the accuracy of our model (DCLNN) compared against SVM. Interestingly, after only 3 epochs, DCLNN almost converged with about $90\%$ accuracy (with final value around $93\%$). The test accuracy also finally converges to $93\%$, whereas SVM classifier (RBF kernel, $C=100$, $\gamma=0.01$) achieves only $85\%$ accuracy. We did not plot F1-score as our class distribution is equal.

\underline{Variation with Channels.} Fig.~\ref{fig:test_acc_vs_chans} shows the variation with channel data. We can see that 3-channel gyroscope data works better (with about $98\%$ accuracy) than 3-channel accelerometer ($93\%$ accuracy) data. Within 3-channel gyroscope, we can notice that axes Y and Z give better performance than axis X. By knowing this, we can discard axis X data and only process axes Y and Z to limit the amount of computation. We can see that axes Y and Z combined can give an accuracy of about $87\%$ accuracy, which could be sufficient in some cases.

\begin{table}[t]
\footnotesize
\caption{Inference Times for AutoEnc on different hardware.}\label{table:desktop-vs-pi}
\vspace{-0.12in}
\begin{center}
\setlength\tabcolsep{1.5pt} %
\begin{tabular}{|p{1.4cm}|p{1.3cm}|p{1.4cm}|p{1.5cm}|p{1.5cm}|}
\hline
\textbf{No. Chans.}                        & \textbf{Desktop (Training)} & \textbf{Desktop (Inference)} & \textbf{RaspberryPi (Inference)} & \textbf{Jetson TX2 (Inference)} \\
\hline
\hline
1 (x/y/z)  & 202 ms                      & 82 ms                        & 312 ms   & 83 ms                  \\
\hline
2 (xy/yz/xz) & 386 ms                      & 87 ms                       & 372 ms   & 92 ms                  \\
\hline
3 (xyz)    & 561 ms                      & 95 ms                       & 484 ms & 101 ms
\\
\hline
\end{tabular}
\vspace{-0.3in}
\end{center}
\end{table}

\underline{Raspberry-Pi and Nvidia Jetson Profling.} The above results are obtained on a desktop computer with Intel QuadCore i7-2600 3.4 GHz processor with 8GB RAM. However, we wanted to know how long it takes to do inference on hardware that can be mounted on a drone and powered using a drone battery. For this purpose, we considered two embedded computing devices---Raspberry Pi 3 Model B and Nvidia Jetson TX2 module, both of which can be mounted on a drone to augment its computing capabilities. The former has a QuadCore 1.2 GHz Broadcom BCM2837 processor with 1GB RAM, the latter has Quadcore 2 GHz ARM processor with 8GB of CUDA-compatible graphic memory. The results are shown in Table~\ref{table:desktop-vs-pi}. The numbers indicate the time taken to do inference on a single window of data (100 samples) with the specified number of channels and the hardware. These numbers show that AutoEnc is amenable for real-time inference on a drone aiding in detection of potentially danger modes. Specifically, we can observe that in the case of Nvidia Jetson TX2, the results are impressive around $100$ ms or less, which means the drone can do inference on the last one second of sensor data every $100$ ms. It is to be noted that, while new sensor data is sampled/generated every 10 ms, it may not be necessary to do the inference at the same speed.

\balance

\section{Conclusion and Future Work}\label{sec:conclusions}
We proposed novel deep architectures to detect and identify the cause of the malfunction. We have shown that the proposed architecture is able to achieve over $90\%$ accuracy for detection and upto 85\% accuracy for identification over experimental data (all-channels combined) and $99\%$ over simulation data (just 3-channels).\\
As future work, we plan to test our model on more crash/attack scenarios e.g., partially broken but functional propellers, cyber attacks, etc. We also plan on testing our models on heterogeneous UAV platforms to assess their generalizability potential.
Later, we will develop real-time decision making techniques to safeguard the drone from the identified misoperation.

\textbf{Acknowledgements:}
We thank PhD student Sriharsha Etigowni for his help in the experiments. This work was supported by the ONR YIP Grant No.~11028418 and by the NSF CNS-1453046.

\balance

\newpage

\bibliographystyle{IEEEtran}\small
\bibliography{bibliography,references}

\end{document}